\documentclass[11pt]{article}

\usepackage{amsmath, amsfonts, bm, amssymb}
\usepackage{amsfonts, graphicx, grffile, float, latexsym, mathtools}
\usepackage{dsfont}
\usepackage{multicol}
\usepackage{comment}
\usepackage{cite}
\usepackage{tikz}
\usetikzlibrary{positioning}
\usetikzlibrary{calc}
\usepackage{wrapfig}
\usepackage{mathrsfs}
\makeatletter
\let\NAT@parse\undefined
\makeatother
\usepackage{amsthm}
\usepackage{bbm}
\usepackage{bm}
\let\labelindent\relax
\makeatletter
\renewcommand*\env@matrix[1][c]{\hskip -\arraycolsep
  \let\@ifnextchar\new@ifnextchar
  \array{*\c@MaxMatrixCols #1}}

\usepackage{Shorthands}
\usepackage{enumitem}

\makeatletter
\let\NAT@parse\undefined
\makeatother
\usepackage{hyperref}
\hypersetup{%
    pdfborder = {0 0 0}
}

\usepackage{cleveref}
\crefname{assumption}{}{}

\newtheoremstyle{named}{}{}{\itshape}{}{\bfseries}{.}{.5em}{\thmnote{#3}}
\theoremstyle{named}

\newcommand{\charis}[1]{{\color{blue}{#1}}}

\newcommand{\bbar}[1]{\bm\bar{#1}}
\newcommand{\barx}{\bbar{x}}
\newcommand{\baru}{\bbar{u}}
\newcommand{\bary}{\bbar{y}}
\newcommand{\barA}{\bbar{A}}
\newcommand{\barB}{\bbar{B}}
\newcommand{\barC}{\bbar{C}}
\newcommand{\barm}{\bbar{m}}
\newcommand{\barn}{\bbar{n}}
\newcommand{\barSigma}{\bbar{\Sigma}}
\newcommand{\barX}{\bbar{\calX}}
\newcommand{\barU}{\bbar{\calU}}
\newcommand{\Acl}{A_{\textup{cl}}}

\usepackage{algorithm} 
\usepackage{algpseudocode} 
\usepackage{authblk}
\newcommand{\tas}[1]{{\color{cyan}{TT: #1}}}
\usepackage{caption}
\renewcommand{\qed}{\hfill\blacksquare}

\usepackage{fullpage}

\title{Layered Multirate Control of Constrained Linear Systems}

\author[1]{Charis Stamouli}
\author[2]{Anastasios Tsiamis}
\author[1]{Manfred Morari}
\author[1]{George J. Pappas}

\affil[1]{University of Pennsylvania}
\affil[2]{ETH Zürich}

\date{}

\begin{document}

\maketitle
\thispagestyle{empty}
\pagestyle{empty}
\captionsetup[figure]{labelfont={bf},labelformat={default},labelsep=period,name={Fig.}}

\begin{abstract}
Layered control architectures have been a standard paradigm for efficiently managing complex constrained systems. A typical architecture consists of: i) a higher layer, where a low-frequency planner controls a simple model of the system, and ii) a lower layer, where a high-frequency tracking controller guides a detailed model of the system toward the output of the higher-layer model. A fundamental problem in this layered architecture is the design of planners and tracking controllers that guarantee both higher- and lower-layer system constraints are satisfied. Toward addressing this problem, we introduce a principled approach for layered multirate control of linear systems subject to output and input constraints. Inspired by discrete-time simulation functions, we propose a streamlined control design that guarantees the lower-layer system tracks the output of the higher-layer system with computable precision. Using this design, we derive conditions and present a method for propagating the constraints of the lower-layer system to the higher-layer system. The propagated constraints are integrated into the design of an arbitrary planner that can handle higher-layer system constraints.  Our framework ensures that the output
constraints of the lower-layer system are satisfied at all high-level time steps, while respecting its input constraints at all low-level time steps. We apply our approach in a scenario of motion planning, highlighting its critical role in ensuring collision avoidance. 
\end{abstract}

\section{Introduction}\label{sec:Introduction}
\begin{figure}[t]
    \centering
    \includegraphics[width=0.65\linewidth]{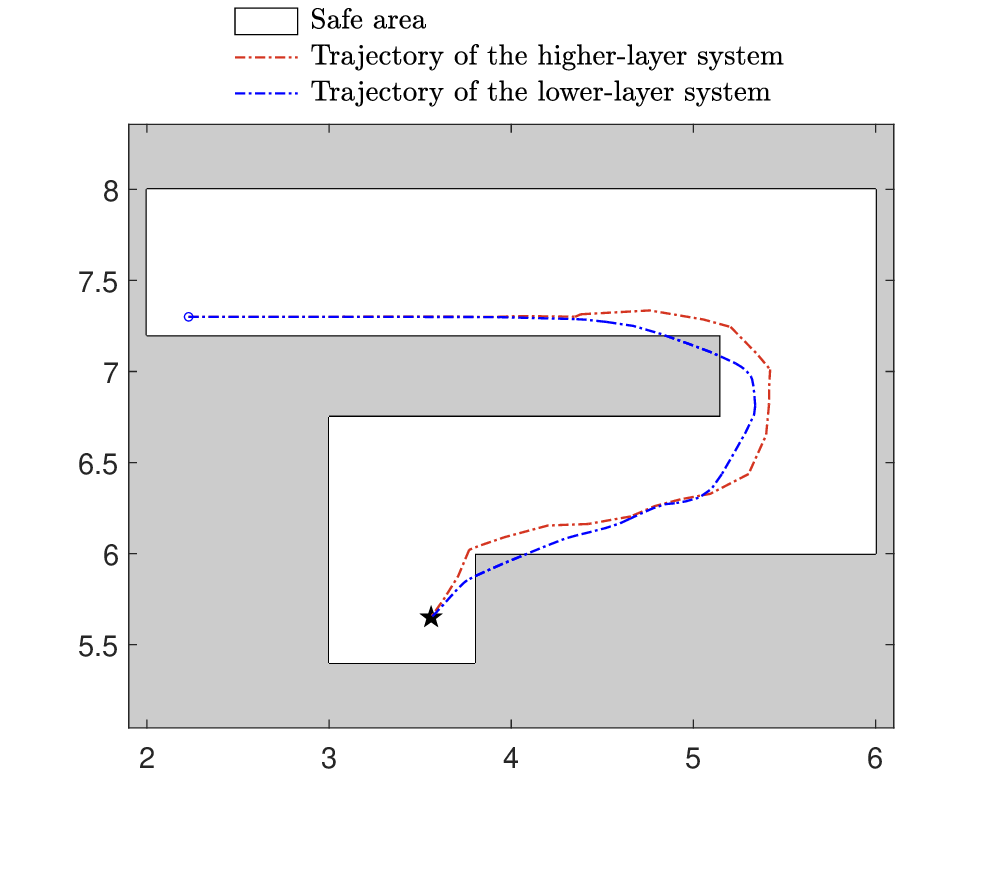}
    \caption{We observe that although the higher-layer system maintains a safe trajectory, the trajectory of the lower-layer system deviates from it sometimes, entering the unsafe area. These deviations arise due to input constraints and additional dynamics in the lower-layer system.}
    \label{fig:failure_plot}
\end{figure}

Layered control architectures have been a standard paradigm for tractable control of complex constrained systems~\cite{matni2024towards}. A typical architecture consists of two layers. The lower layer uses a detailed model of the system, while the higher layer uses a simpler model of the system. The control inputs of the higher-layer system are provided by a low-frequency planner that accounts for complex higher-layer constraints. A high-frequency tracking controller refines the inputs from the planner to steer the lower-layer system toward the trajectory of the higher-layer system. 
%
A fundamental problem in this layered architecture is the design of planners and tracking controllers that ensure the satisfaction of both higher- and lower-layer system constraints. More specifically, the presence of actuation limits and additional dynamics in the lower-layer system can cause its trajectories to deviate from those of the higher-layer system. These deviations can potentially lead the lower-layer system to unsafe regions (see Fig.~\ref{fig:failure_plot}) and thus cannot be neglected. 

\begin{figure}[tbh]
\begin{center}
    \begin{tikzpicture}
        \node[draw=black, thick, minimum width=4cm, minimum height=2cm] (top) at (0,2) {
            \begin{minipage}{6.5cm}
                \centering
                \vspace{-0.3cm}
                \begin{align*}
                    &\hspace{1.2cm}\text{Low frequency $1/T_H$}\\[5pt]                    &\hspace{.6cm}\barSigma:\left\{
                        \begin{array}{lcr}
                            \dot\barx(t)=\barA\barx(t)+\barB\baru(t) \\
                            \bary(t) = \barC\barx(t)
                        \end{array}
                        \right.\\[5pt]
                        &\barx(t)\in\barX,\;\baru(t)\in\barU,\; t=0,T_H,2T_H,\ldots 
                \end{align*}
            \end{minipage}
        };
        
        \node[draw=black, thick, minimum width=4cm, minimum height=2cm] (bottom) at (0, -1.5) {
            \begin{minipage}{6.5cm}
                \centering
                \vspace{-0.3cm}
                \begin{align*}
                    &\hspace{1.1cm}\text{High frequency $1/T_L$}\\[5pt]                    &\hspace{.6cm}\Sigma:\left\{
                        \begin{array}{lcr}
                            \dot x(t)=Ax(t)+Bu(t) \\
                            y(t) = Cx(t)
                        \end{array}
                        \right.\\[5pt]
                        &y(t)\in\mathcal{Y},\; u(t)\in\mathcal{U},\; t=0,T_L,2T_L\ldots
                \end{align*}
            \end{minipage}
        };
        
        \draw[->, thick] (top.south) -- (bottom.north) node[midway, right] {$\barx(t),\baru(t)$};
    \end{tikzpicture}
\end{center}
 \caption{Two-layer multirate control architecture for linear systems with output and input constraints. $\Sigma$ is the system that we want to control, while $\barSigma$ provides a simplified description of $\Sigma$'s dynamics.}
\label{fig:problem_architecture}
\end{figure}
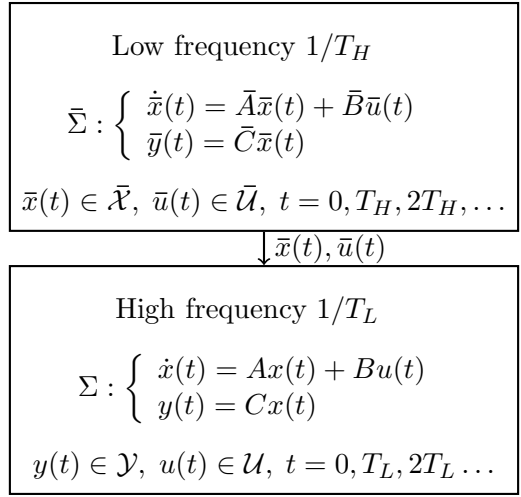

In this paper, we propose a principled layered control approach for linear systems subject to output and input constraints (see Fig.~\ref{fig:problem_architecture}). We assume we are given a planner that can handle higher-layer system constraints and integrate the model mismatch and the lower-layer system constraints into the architectural design. Our approach provides a joint design of tracking controllers and planning constraints that account for both higher- and lower-level system constraints, drawing inspiration from simulation functions \cite{Girard2007,Girard2009}. 
Simulation functions are control Lyapunov-like functions that describe how a state trajectory of the higher-layer system can be mapped to a state trajectory of the lower-layer system such that the distance between the corresponding output trajectories remains within computable bounds. Our contributions are the following:
\begin{itemize}
\item  We provide a systematic approach for designing tracking controllers that ensure output tracking with computable precision at all low-level time steps.
\item Employing our tracking control design, we derive conditions and present a method for propagating the lower-layer system constraints to the higher-layer system.
\item We prove that the output constraints of the lower-layer system are satisfied at every high-level time step, while its input constraints are satisfied at every low-level time step. Our framework provides a streamlined and principled approach to layered multirate control of linear systems subject to output and input constraints.
\end{itemize}

We apply our framework to navigate a robot in a maze-like environment and observe that higher sampling frequencies for the tracking control implementation result in improved tracking precision. Furthermore, our approach ensures the avoidance of the unsafe scenario depicted in Fig.~\ref{fig:failure_plot}, while the tracking precision of our controller is observed to provide a quite tight upper bound for the output distance of the systems during the mission.

All proofs can be found in Appendix~\ref{app:Proofs}.
\medskip

\noindent\textbf{Related Work.} The problem of layered control design has been extensively studied in recent years, with a plethora of architectures proposed in the literature. Implementations have been presented in multirate \cite{grandia2021multi,rosolia2020multi,rosolia2022unified}, single-rate \cite{srikanthan2023data,srikanthan2024closed}, and continuous-time \cite{cohen2024safety,herbert2017fastrack,singh2020robust, yin2020optimization,schweidel2022safe} settings. 

The above works primarily address nonlinear lower-layer systems with state constraints, and propose different controllers for \emph{state tracking}. For example, controllers are designed based on quadratic programming in \cite{rosolia2020multi,rosolia2022unified}, Hamilton-Jacobi reachability analysis in \cite{herbert2017fastrack}, and SOS programming in \cite{singh2020robust,yin2020optimization,schweidel2022safe}. In contrast, we focus on obtaining a streamlined method for \emph{output tracking} tailored to linear lower-layer systems with output constraints. Our setting provides increased flexibility by evaluating the distance between the lower- and higher-layer systems based on their outputs, which constitute a transformed lower-dimensional representation of their states. We also note that input constraints are considered in some of the aforementioned works, specifically in \cite{rosolia2020multi,rosolia2022unified,srikanthan2024closed,yin2020optimization,schweidel2022safe}.

An alternative layering approach for linear systems with state and input constraints is that of hierarchical model predictive control (MPC) \cite{scattolini2007hierarchical,Raghuraman2020,farina2018hierarchical}.
In hierarchical MPC, each layer runs an MPC at a different sampling frequency, with lower layers refining the control actions generated in higher layers. In contrast, our approach relies on an arbitrary planner for constraint satisfaction and a linear tracking controller at the lower layer of the architecture.

A layered architecture similar to ours is studied in \cite{barcelliy2010hierarchical}. Therein, the reference signals to the lower layer do not necessarily follow higher-layer dynamics, and
constraint propagation relies on a predefined stabilizing controller for the lower-layer system. Reference governors~\cite{garone2017reference} follow similar principles. Differently, in our work, simulation functions allow us to integrate higher-layer dynamics into both the constraint propagation and the design of tracking controllers.

Control designs based on simulation functions have been presented in both continuous-time \cite{Girard2007,Girard2009,fainekos2007hierarchical,Fainekos2009} and discrete-time \cite{Lavaei2017,zhong2024hierarchical} settings. The approaches in \cite{Girard2007,Lavaei2017,zhong2024hierarchical} focus on unconstrained linear systems, whereas the framework in \cite{fainekos2007hierarchical,Fainekos2009} ensures robust satisfaction of temporal logic specifications for robots with second-order dynamics. In contrast, we focus on linear systems with output and input constraints and provide a streamlined layered multirate control design with guarantees of output and input constraint satisfaction. We note that simulation functions for linear systems with input constraints are provided in \cite{girard2005approximate}, but corresponding controllers are not derived by the authors.

Although our work focuses on layered approaches, there are multiple alternatives that do not necessarily follow such an explicit separation---see the recent survey~\cite{kohler2024analysis}. 
\medskip 

\noindent\textbf{Notation.} The norm $\norm{\cdot}$ is the Euclidean norm whenever it is applied to vectors and the spectral norm whenever it is applied to matrices. Moreover, $\mathbb{I}_d$ is the $d\times d$ identity matrix and $\calS^d$ is the unit sphere in $\setR^{d}$. Given two sets $\calA,\calB\subseteq\setR^d$, we use $\calA+\calB$ to denote their Minkowski sum and $\calA-\calB$ to denote their Minkowski difference. If $\calA$ is a subset of $\setR^d$ and $D$ is a matrix in $\setR^{d'\times d}$, the set $D\calA:=\{Da: a\in\calA\}$ is the image of $\calA$ under $D$.

\section{Problem Formulation}\label{sec:Problem Formulation}
Consider the layered control architecture depicted in Fig.~\ref{fig:problem_architecture}. The higher layer involves a linear system $\barSigma$ with state $\barx(t)\in\setR^{\barn}$, input $\baru(t)\in\setR^{\barm}$, and output $\bary(t)\in\setR^p$, which is controlled at a given sampling frequency $1/T_H$. We assume that $\barSigma$ is subject to state constraints $\barx(t)\in\barX$ and input constraints $\baru(t)\in\barU$, where $\barX\subseteq\setR^{\barn}$ is closed and $\barU\subseteq\setR^{\barm}$ is compact, for all $t=hT_H$, $h\in\setN$. In the lower layer, a linear system $\Sigma$ with state $x(t)\in\setR^{n}$, input $u(t)\in\setR^{m}$, and output $y(t)\in\setR^p$, where $p\leq n$, is controlled at a higher sampling frequency $1/T_L$.\footnote{We assume that $T_H$ and $T_L$ are such that $T_H/T_L\in\setN_+$.\label{fn1:rates_ratio}} We assume that $\Sigma$ is subject to output constraints $y(t)\in\calY$ and input constraints $u(t)\in\calU$, where $\calY\subseteq\setR^p$ is closed and $\calU\subseteq\setR^m$ is compact, for all $t=\ell T_L$, $\ell\in\setN$. 

Suppose that the states of  $\barSigma$ and $\Sigma$ can be measured directly and satisfy $(\barx(0),x(0))\in\calI$ at time $0$, where:
\[
\calI\subseteq\{(\barx,x)\in\setR^{\barn}\times\setR^n:\barC\barx=Cx,\barx\in\barX,Cx\in\calY\}
\]
is compact. We highlight that $\Sigma$ and $\barSigma$ have the same output dimension, but their state and input dimensions may differ.
In practice, $\barSigma$ typically provides a simplified representation of $\Sigma$'s dynamics, and hence, we assume that $\barn\leq n$. As an example, $\Sigma$ may capture the dynamics of a mobile robot, while $\barSigma$ represents its kinematics. A more concrete example for this architecture is given in Section~\ref{sec:Case Study: Navigation of a Mobile Robot}.

Assume we are given a planner that maps the state $\barx(t)$ of $\barSigma$ to a control input $\baru(t)$ and can integrate state and input constraints for $\barSigma$. For example, consider a model predictive controller or a Rapidly-exploring Random Tree Star variant \cite{Webb2012KinodynamicRO}. The input $\baru(t)$ is typically determined by optimizing a performance criterion, such as a tracking objective.  The state $\barx(t)$ and input $\baru(t)$ of system $\barSigma$ are passed to the lower layer of Fig.~\ref{fig:problem_architecture}, where a tracking controller aims to steer the outputs $y(t)$ of $\Sigma$ toward the outputs $\bary(t)$ of $\barSigma$. A main challenge in this layered architecture is the design of planners and tracking controllers such that all constraints are guaranteed to be satisfied. Consider, for example, the scenario depicted in Fig.~\ref{fig:failure_plot}, where deviations of the outputs $y(t)$ from the outputs $\bary(t)$ due to model mismatch and input constraints of $\Sigma$ lead the output $y(t)$ of $\Sigma$ to unsafe regions.

To ensure the satisfaction of both higher- and lower-layer system constraints, we consider imposing potentially tighter constraints for $\barSigma$ in the planner. Specifically, let $\barX_p\subseteq\barX$ and $\barU_p\subseteq\barU$ denote the planning constraints for the state and input of $\barSigma$, respectively. By design, such constraint sets guarantee the satisfaction of the given constraints for $\barSigma$. Our goal is to jointly design tracking controllers with guaranteed precision and suitable planning constraint sets $\barX_p$ and $\barU_p$ such that the lower-layer system constraints are also ensured to be satisfied. Our problem is formalized in the following statement.

\begin{problem}[Layered Multirate Control of Constrained Linear Systems]\label{problem}
Consider the layered multirate control architecture depicted in Fig.~\ref{fig:problem_architecture}. Consider a mission of total time $T$ and suppose that $T/T_H\in\setN_+$.\footref{fn1:rates_ratio} Assume we are given an arbitrary planner that can integrate constraint sets $\barX_p\subseteq\barX$ and $\barU_p\subseteq\barU$ for the state and input of $\barSigma$, respectively. Design a tracking controller for $\Sigma$ and planning constraint sets $\barX_p$ and $\barU_p$ such that, for any pair of initial states $(\barx(0),x(0))\in\calI$ with $\barx(0)\in\barX_p$:
\begin{enumerate}[label=\roman*)]
    \item The outputs of the closed-loop systems $\barSigma$ and $\Sigma$ satisfy;
    \begin{equation}\label{eq:precision_guarantee}
        \norm{\bary(t)-y(t)}\leq\varepsilon, \forall t=0,T_L,\ldots,T,
    \end{equation}
    where $\varepsilon>0$ is a computable bound;
    \item The closed-loop system $\Sigma$ satisfies the lower-layer output and input constraints.
\end{enumerate}
\end{problem}


In the next section, we derive a tracking controller that allows system $\Sigma$ to track any output trajectory of $\barSigma$ with computable precision as in \eqref{eq:precision_guarantee}. In Section~\ref{sec:Hierarchical Constraint Propagation}, the resulting controller and its precision are leveraged in the design of constraint sets $\barX_p$ and $\barU_p$ that provide guarantees of output and input constraint satisfaction for system $\Sigma$.

\section{Output Tracking with Guaranteed Precision}\label{sec:Tracking using Approximate Simulation}
In this section, we present a control framework that enables the lower-layer system to track any output trajectory of the higher-layer system with computable precision. Our control design leverages simulation functions for discrete-time systems \cite{Ma2015,Lavaei2017} to establish a guaranteed precision for all $t=\ell T_L$, $\ell\in\setN$ (see Subsection~\ref{subsec:Guaranteed Tracking Error Bounds via Discrete-Time Simulation Functions}). A systematic method of computing discrete-time simulation functions and corresponding tracking controllers is provided in Subsection~\ref{subsec:Computation of Tracking Controllers based on Discrete-Time Simulation Functions}. 

\subsection{Guaranteed Tracking Precision via Discrete-Time Simulation Functions}\label{subsec:Guaranteed Tracking Error Bounds via Discrete-Time Simulation Functions}
In this subsection, we introduce a notion of discrete-time simulation functions and corresponding tracking controllers with guaranteed precision. Our definitions result from adapting those in \cite{Girard2009} to discrete-time systems.

Consider the exact discretization of $\Sigma$ and $\barSigma$  with zero-order hold inputs \cite{chen1984linear} at the lower-layer sampling frequency $1/T_L$, given by:
\begin{equation*}
    \Sigma_L:\left\{
            \begin{array}{lcr}
               x_{\ell+1}=A_Lx_\ell+B_Lu_\ell \\
                y_\ell = Cx_\ell
            \end{array}
            \right.     
\end{equation*}
and:
\begin{equation*}
    \barSigma_L:\left\{
            \begin{array}{lcr}
               \barx_{\ell+1}=\barA_L\barx_\ell+\barB_L\baru_\ell \\
                \bary_\ell = \barC\barx_\ell
            \end{array}
            \right.,            
\end{equation*}
respectively. Intuitively, a simulation function of $\barSigma_L$ by $\Sigma_{L}$ is a function over their state spaces describing how a state trajectory of $\barSigma_{L}$ can be transformed into a state trajectory of $\Sigma_{L}$ such that the distance between the respective output trajectories remains within computable bounds. Simulation functions of $\barSigma_L$ by $\Sigma_L$ and corresponding tracking controllers are formally defined next.

\begin{definition}[Discrete-Time Simulation Functions and Corresponding Tracking Controllers]\label{def:Discrete-Time Simulation Functions and Associated Tracking Controllers}
A continuous function $V:\setR^{\barn}\times\setR^{n}\to\setR_+$ is a simulation function of $\barSigma_L$ by $\Sigma_L$ and a function $u_L:\setR^{\barm}\times\setR^{\barn}\times\setR^n\to\setR^m$ is a corresponding tracking controller if there exists $\gamma>0$ such that for all
$(\barx_\ell,x_\ell)\in\setR^{\barn}\times\setR^n$:
\begin{align}
\label{eq:SF_condition_1}
    V&(\barx_\ell,x_\ell)\geq \norm{\barC\barx_\ell-Cx_\ell}^2
\end{align}
and for all $\baru_\ell\in\setR^{\barm}$ satisfying $\gamma^2\norm{\baru_\ell}^2<V(\barx_\ell,x_\ell)$:
\begin{equation}
\label{eq:SF_condition_2}
    V(\barx_{\ell+1},x_{\ell+1})-V(\barx_\ell,x_\ell)<0,
\end{equation}
where $\barx_{\ell+1}=\barA_L\barx_\ell+\barB_L\baru_\ell$ and $x_{\ell+1}=A_Lx_\ell+B_Lu_L(\baru_\ell,\barx_\ell,x_\ell)$. 
\end{definition}
Condition \eqref{eq:SF_condition_1} implies that a simulation function $V(\cdot,\cdot)$ provides an upper bound for the squared output distance of systems $\Sigma_L$ and $\barSigma_L$, given any pair of their states. Moreover, a corresponding controller $u_L(\cdot,\cdot,\cdot)$ guarantees the decrease of $V(\cdot,\cdot)$ along the systems' evolution whenever $V(\barx_\ell,x_\ell)>\gamma^2\norm{\baru_\ell}^2$ (see \eqref{eq:SF_condition_2}). By applying the controller $u_L(\cdot,\cdot,\cdot)$, we can ensure that system $\Sigma_L$ is able to track any output trajectory of system $\barSigma_L$ with computable precision.

\begin{lemma}[Tracking Precision Guarantee]\label{lem:tracking_precision_guarantee}
Let $V:\setR^{\barn}\times\setR^{n}\to\setR_+$ be a simulation function of $\barSigma_L$ by $\Sigma_L$ and $u_L:\setR^{\barm}\times\setR^{\barn}\times\setR^n\to\setR^m$ be a corresponding tracking controller. Fix a pair of initial states $(\barx_0,x_0)\in\calI$ and consider inputs $\baru_\ell\in\barU_p$ for $\barSigma_L$, where $\barU_p$ is a compact set. Let $\baru_{\max}>0$ be such that $\max_{\baru\in\barU_p}\norm{\baru}\leq\baru_{\max}$. Consider the states $x_\ell$ of $\Sigma_L$ resulting from the dynamics equation:
\begin{equation*}
    x_{\ell+1} = A_Lx_\ell+B_Lu_L(\baru_\ell,\barx_\ell,x_\ell),
\end{equation*}
and let $y_\ell$ denote the associated outputs. Then, for all $\ell\in\setN$, we have:
\begin{equation}\label{eq:SF_guarantee}
    \norm{\bary_\ell-y_\ell}\leq\max\left\{\sqrt{V(\barx_0,x_0)},\gamma\baru_{\max}\right\}.
\end{equation}
\end{lemma}
The bound on the right-hand side of \eqref{eq:SF_guarantee} depends on the initial states of the systems and an upper bound $\baru_{\max}$ for the maximum norm of inputs allowed in $\barU_p$. The set $\barU_p$ can be thought of as the planning constraint set for the input of the higher-layer system (see Problem~\ref{problem}). We define the uniform tracking error bound over all initial states as: 
\begin{equation}\label{eq:precision}
\varepsilon=\max\left\{\max_{(\barx_0,x_0)\in\calI}\sqrt{V(\barx_0,x_0)},\gamma\baru_{\max}\right\}.
\end{equation}
Then, inequality \eqref{eq:SF_guarantee} implies that for any pair of initial states $(\barx_0,x_0)\in\calI$ and any input trajectory of $\barSigma_L$ allowed by $\barU_p$, the controller $u_L(\cdot,\cdot,\cdot)$ ensures the output of $\Sigma_L$ remains within $\varepsilon$ of the output of $\barSigma_L$. Throughout the paper, we refer to $\varepsilon$ as the tracking precision of $\barSigma_L$ by $\Sigma_L$ corresponding to the controller $u_L(\cdot,\cdot,\cdot)$. In applications of interest, the first term in the maximum of \eqref{eq:precision} is usually negligible, leading to $\varepsilon=\gamma\baru_{\max}$ (see example in Section~\ref{sec:Case Study: Navigation of a Mobile Robot}).

\subsection{Computation of Tracking Controllers based on Discrete-Time Simulation Functions}\label{subsec:Computation of Tracking Controllers based on Discrete-Time Simulation Functions}
In this subsection, we present a systematic method of designing discrete-time simulation functions of $\barSigma_L$ by $\Sigma_L$ and corresponding controllers $u_L(\cdot,\cdot,\cdot)$. Our approach is inspired by \cite{Girard2009}, where simulation functions and respective controllers are derived for continuous-time linear systems.  

Before presenting our main result, we introduce an assumption about systems $\Sigma_L$ and $\barSigma_L$, along with a lemma that is employed in its derivation.

\begin{assumption}\label{ass:P_conditions}
System $\Sigma_L$ is stabilizable. Moreover, there exist matrices $P\in\setR^{n\times\barn}$ and $Q\in\setR^{m\times\barn}$ that satisfy: 
\begin{align}
\label{eq:P_condition}
&CP=\barC, \\
\label{eq:P_Q_condition}
&P\barA_L=A_LP+B_LQ.
\end{align}
\end{assumption}

Stabilizability of $\Sigma_L$ is a minimal assumption. If $\barSigma_L$ does not have any input (i.e., $B_L=0$), our tracking problem becomes similar to the classical regulator problem \cite{Wonham1979}, for which equations
\eqref{eq:P_condition} and \eqref{eq:P_Q_condition} are key ingredients (see \cite[Remark 4.4]{Girard2007} for details). We note in passing that analogous conditions are used in prior work \cite{Lavaei2017,Girard2009}, but an approach similar to \cite{zhong2024hierarchical} could be used to remove condition \eqref{eq:P_Q_condition}.

\begin{lemma}\label{lem:M_K_conditions}
There exists a matrix $K\in\setR^{m\times n}$, a positive definite matrix $M\in\setR^{n\times n}$, and a scalar $\lambda\in(0,1/2)$ such that the following matrix
inequalities hold:
\begin{align}
    \label{eq:M_condition}
    &M\succeq C^{\intercal}C, \\
    \label{eq:M_K_condition}
    &(A_L+B_LK)^{\intercal}M(A_L+B_LK)-M\preceq-2\lambda M.
\end{align}
\end{lemma}

Lemma~\ref{lem:M_K_conditions} relies on our stabilizability assumption for system $\Sigma_L$. A simple method to compute $K$, $M$, and $\lambda$ satisfying the conditions of Lemma~\ref{lem:M_K_conditions} is provided in its proof.  Appendix~\ref{app:Computation of M and K using Semidefinite Programming} presents an alternative computational approach based on semidefinite programming, which leads to tighter tracking precision. 

The following theorem provides a discrete-time version of \cite[Theorem 2]{Girard2009}. 

\begin{theorem}[Output Tracking using Discrete-Time Simulation Functions]\label{theorem:Tracking_using_Discrete-Time_Simulation_Functions}
Consider matrices $K\in\setR^{m\times n}$, $M\in\setR^{n\times n}$ and a scalar $\lambda\in(0,1/2)$ that satisfy the conditions of Lemma~\ref{lem:M_K_conditions}. Moreover, assume there exist $P\in\setR^{n\times\barn}$ and $Q\in\setR^{m\times\barn}$ that satisfy the conditions of Assumption~\ref{ass:P_conditions}. Then, the function $V:\setR^{\barn}\times\setR^{n}\to\setR_+$ defined by:
\begin{equation}\label{eq:simulation_function}
    V(\barx_\ell,x_\ell)=(x_\ell-P\barx_\ell)^{\intercal}M(x_\ell-P\barx_\ell)
\end{equation}
is a simulation function of $\barSigma_L$ by $\Sigma_L$ and a corresponding tracking controller is given by:
\begin{equation}\label{eq:tracking_controller}
    u_L(\baru_\ell,\barx_\ell,x_\ell) = R\baru_\ell+Q\barx_\ell+K(x_\ell-P\barx_\ell),
\end{equation}
where $R\in\setR^{m\times\barm}$ is an arbitrary matrix. The scalar $\gamma$ corresponding to the simulation function is given by: 
\begin{equation*}
    \gamma = \frac{\sqrt{1-\lambda}\,\norm{M^{1/2}(B_LR-P\barB_L)}}{\lambda}
\end{equation*}
and is minimal for $R=(B_L^{\intercal}MB_L)^{-1}B_L^{\intercal}MP\barB_L$.
\end{theorem}

Notice that the first two terms in \eqref{eq:tracking_controller} guide the higher-layer output toward a desired value, while the last term ensures that the lower-layer state remains close to the higher-layer state lifted to the image of the matrix $P$. From Lemma~\ref{lem:tracking_precision_guarantee} we can conclude that tracking controllers of the form \eqref{eq:tracking_controller} guarantee that the output trajectory of $\Sigma_L$ remains $\varepsilon$-close to \textit{any} output trajectory of $\barSigma_L$.

\begin{remark}\label{rem:comparison_to_Lavaei}
Discrete-time simulation functions and corresponding tracking controllers for linear systems were presented in \cite{Lavaei2017,zhong2024hierarchical}. The approaches in \cite{Lavaei2017,zhong2024hierarchical} consider matrices $M$, $K$, and a scalar $\kappa$ that satisfy linear matrix inequalities (LMIs) similar to \eqref{eq:M_condition} and \eqref{eq:M_K_condition}. However, it is unclear when the LMIs employed in \cite{Lavaei2017,zhong2024hierarchical} admit a solution. In contrast, we prove the existence of matrices $M$, $K$, and a scalar $\lambda$ that satisfy the LMIs \eqref{eq:M_condition} and \eqref{eq:M_K_condition}, given any stabilizable system $\Sigma_L$. Beyond that, the proof of Lemma~\ref{lem:M_K_conditions} provides a systematic method of computing the aforementioned quantities, while an alternative computational approach that leads to tighter tracking precision is provided in Appendix~\ref{app:Computation of M and K using Semidefinite Programming}.
\end{remark}

Our $\varepsilon$-tracking guarantee is essential for propagating the output and input constraints of the lower-layer system to the higher-layer system, as detailed in the following section.  

\section{Constraint Propagation}\label{sec:Hierarchical Constraint Propagation}
In the previous section, we presented a systematic control design that guarantees the output of system $\Sigma_L$ remains $\varepsilon$-close to the output of system $\barSigma_L$. By leveraging our design and its precision $\varepsilon$, we can propagate the constraints of the lower-layer system to the higher-layer system and obtain tightened constraint sets $\barX_p\subseteq\barX$ and $\barU_p\subseteq\barU$ for the planner that controls the higher-layer system (see Problem~\ref{problem}).

Constraint propagation can be achieved by shrinking the constraint sets $\calY$ and $\calU$ of the lower-layer system based on the tracking control design in \eqref{eq:tracking_controller} and the associated precision $\varepsilon$. In the following proposition, we characterize conditions for the sets $\barX_p$ and $\barU_p$ that provide guarantees of output and input constraint satisfaction for system $\Sigma$.

\begin{figure}[tbh]
\begin{center}
\begin{tikzpicture}[node distance=1.5cm]
    \node[draw, minimum width=6.5cm, minimum height=0.6cm, align=center] (planner) {\texttt{Planner}};
    \node[draw, minimum width=3cm, minimum height=1.5cm, align=center, below=1.2cm of planner] (system1) {
        \hspace{0.cm}
        System $\barSigma$
    };
    \node[draw, minimum width=6.5cm, minimum height=0.6cm, align=center, below=1.5cm of system1] (controller) {\textcolor{violet}{Tracking controller \( u_L \)}};
    \node[draw, minimum width=3cm, minimum height=1.5cm, align=center, below=.8cm of controller] (system2) {
        \hspace{0.cm}
        System $\Sigma$
    };
     \draw[->] (planner.south) -- node[left, yshift=5pt]  {Low frequency $1/T_H$\hspace*{0.7cm}} node[right, yshift=5pt] {$\baru\in\textcolor{violet}{\barU_p}\subseteq\barU$} (system1.north);
    \draw[->] (system1.south) -- node[right, yshift=8pt] {$\barx\in\textcolor{violet}{\barX_p}\subseteq\barX$} (controller.north);
    \draw[->] (controller.south) -- node[left, yshift=0pt]  {High frequency $1/T_L$\hspace*{0.7cm}} node[right, yshift=0pt]  {$u\in\calU$} (system2.north);
    \draw[->] ($(planner.south)!0.7!(system1.north)$) --++ (-2.5,0) --++ (0,-3.38);
    \draw[->] ($(system1.south)!0.6!(controller.north)$) --++ (3.5,0) --++ (0,4.55) --++ (-3.5,0) --++(0,-0.32);
    \draw[->] (system2.south) --node[right, yshift=0pt]  {$x$}++(0,-0.6) --++(3.5,0) --++(0,3.82) --++ (-1.5,0) --++(0,-0.31);
    \draw[dashed,->] (system1.east) -- ++(2.4,0) node[above, xshift=-1.4cm] {$\bary\in\barC\textcolor{violet}{\barX_p}$};
    \draw[dashed,->] (system2.east) -- ++(2.4,0) node[above, xshift=-1.53cm] {$y\in\calY$};
\end{tikzpicture}
\end{center}
   \caption{We propose a joint design of tracking controllers $u_L$ and planning constraint sets $\barX_p$ and $\barU_p$. The sets $\barX_p$ and $\barU_p$ result from propagating the constraints $\calY$ and $\calU$ of the lower-layer system to the higher-layer system and combining them with the given constraint sets $\barX$ and $\barU$ of the latter.}
   \label{fig:hier_control_architecture}
\end{figure}
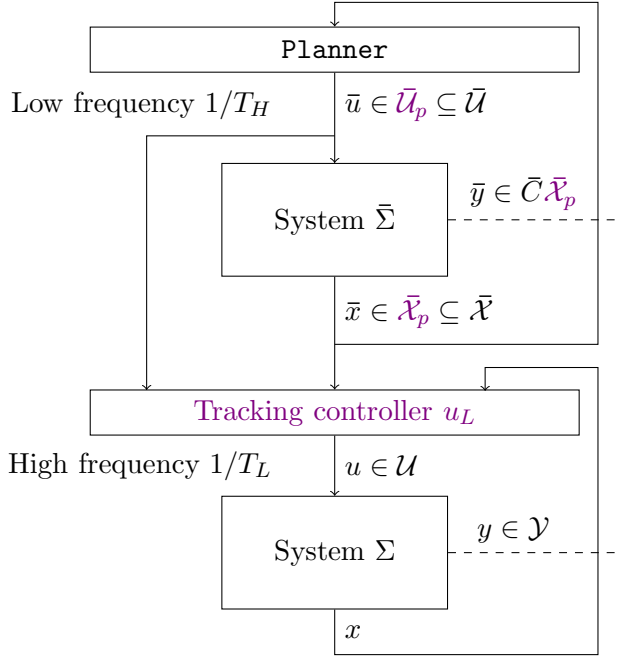

\begin{proposition}[Conditions for Constraint Propagation]\label{prop:Hierarchical Constraint Propagation}
Consider the layered multirate control architecture depicted in Fig.~\ref{fig:hier_control_architecture}. Let $\texttt{Planner}:\setR^{\barn}\to\setR^{\barm}$ denote a planner parametrized by constraint sets $\barX_p\subseteq\barX$ and $\barU_p\subseteq\barU$ for the state and input of $\barSigma$, respectively. Let the tracking controller $u_L:\setR^{\barm}\times\setR^{\barn}\times\setR^n\to\setR^m$ be of the form \eqref{eq:tracking_controller}
and let the tracking precision $\varepsilon$ be as in \eqref{eq:precision} with $\baru_{\max}\geq\max_{\baru\in\barU_p}\norm{\baru}$.  Assume that $\barX_p$ and $\barU_p$ are nonempty sets and satisfy the following conditions:
\begin{align}
    \label{eq:constraints_condition_1}
    &\barC\barX_p\subseteq\calY-\varepsilon\calS^p,\\
    \label{eq:constraints_condition_2}
    &R\,\barU_p+Q\barX_p\subseteq\calU-\varepsilon KM^{-1/2}\calS^n.
\end{align}
Then, for any pair of initial states $(\barx(0),x(0))\in\calI$ with $\barx(0)\in\barX_p$, the closed-loop system $\Sigma$ is guaranteed to satisfy:
\begin{align}
    \label{eq:safety_guarantee}
    &y(t)\in\calY, \forall t=0,T_H,\ldots,T,\\
    \label{eq:input_constraint_guarantee}
    &u(t)\in\calU, \forall t=0,T_H,\ldots,T.    
\end{align}
Moreover, if $Q=0$, we have:
\begin{align}
    \label{eq:input_constraint_guarantee2}
    u(t)\in\calU, \forall t=0,T_L,\ldots,T.    
\end{align}
\end{proposition}

Condition \eqref{eq:constraints_condition_1} requires that the outputs of the higher-layer system satisfy the output constraints of the lower-layer system by a margin of $\varepsilon$. Since $\Sigma_L$ remains $\varepsilon$-close to the output trajectory of $\barSigma_L$, this condition suffices to ensure the outputs constraints in \eqref{eq:safety_guarantee}. Condition \eqref{eq:constraints_condition_2} ensures that the constrained control inputs and states of the higher-layer system are mapped to feasible inputs $u(t)\in\calU$ for the lower-layer system through the control law in \eqref{eq:tracking_controller}. 

\begin{remark}\label{rem:safety_at_low_level_time_steps}
The guarantees \eqref{eq:safety_guarantee} and \eqref{eq:input_constraint_guarantee} ensure that the output and input constraints of $\Sigma$ are satisfied at every high-level time step, whereas \eqref{eq:input_constraint_guarantee2} ensures that its input constraints are satisfied at every low-level time step if $Q=0$ (see, e.g., the case study in Section~\ref{sec:Case Study: Navigation of a Mobile Robot}). To guarantee satisfaction of the output and input constraints for any matrix $Q$ and all low-level time steps, we could consider additional planning constraints that limit the rate of variation of the states $\barx(t)$, employing ideas from \cite{barcelliy2010hierarchical}. A thorough analysis of this extension is reserved for future work.
\end{remark}

The applicability of our framework relies on our capability of deriving sets $\barX_p$ and $\barU_p$ that satisfy the conditions of~\Cref{prop:Hierarchical Constraint Propagation}. For simplicity, we consider polyhedral constraint sets for the lower-layer system, given by: 
\begin{align*}
    &\calY=\left\{y\in\setR^p: F_yy\leq f_y\right\},\\
    &\calU=\left\{u\in\setR^m: F_uu\leq f_u\right\},
\end{align*}
where $F_y\in\setR^{d_y\times p}$, $f_y\in\setR^{d_y}$, $F_u\in\setR^{d_u\times m}$, and $f_u\in\setR^{d_u}$. Let $F_{y,j}$, $F_{u,j}$, and $G_{u,j}$ be the $j$-th row of the matrices $F_y$, $F_u$, and $G_u := F_uKM^{-1/2}$, respectively. Moreover, let $\tilde{f}_y=(\norm{F_{y,1}},\ldots,\norm{F_{y,d_y}})$, $\tilde{f}_u=(\norm{F_{u,1}}_1,\ldots,\norm{F_{u,d_u}}_1)$, and $\tilde{g}_u=(\norm{G_{u,1}},\ldots,\norm{G_{u,d_u}})$. We define the sets $\barX_p$ and $\barU_p$ as follows:
\begin{align}
    \label{eq:barX_polyhedral}
    \barX_p&=\{\barx\in\barX: F_y\barC\barx\leq f_y-\varepsilon\tilde{f}_y,\norm{Q\barx}_{\infty}\leq\delta\},\\
    \label{eq:barU_polyhedral}
    \barU_p&=\{\baru\in\barU: F_uR\baru\leq f_u-\delta\tilde{f}_u-\varepsilon\tilde{g}_u\},
\end{align}
where $\delta\geq0$ is a hyperparameter. Assume there exist $\baru_{\max}>0$ and $\delta\geq0$ such that the above sets are nonempty and satisfy $\baru_{\max}\geq\max_{\baru\in\barU_p}\norm{\baru}$. Then, we can show that the conditions \eqref{eq:constraints_condition_1} and \eqref{eq:constraints_condition_2} are satisfied (see the proof in Appendix~\ref{app:Proofs}). The hyperparameters $\baru_{\max}$ and $\delta$ can be chosen through trial and error so that $\barX_p$ and $\barU_p$ are nonempty sets and $\baru_{\max}\geq\max_{\baru\in\barU_p}\norm{\baru}$. 
In the special case where $Q=0$, we can pick $\delta=0$ and remove the constraint $\norm{Q\barx}_{\infty}\leq\delta$ from \eqref{eq:barX_polyhedral}.

\begin{remark}\label{rem:derivation_of_planning_constraint_sets}
In the case of general convex constraint sets $\calY$ and $\calU$, we can apply the above derivation method to a polyhedral inner approximation of these sets. More generally, if the constraint sets $\calY$ and $\calU$ are nonconvex but can be written as a union of convex sets, the same approach can be employed for polyhedral inner approximations of the individual convex sets. An example of motion planning in a nonconvex safe area $\calY$ is presented in Section~\ref{sec:Case Study: Navigation of a Mobile Robot}. 
\end{remark}

Next, we combine our results from the previous and current sections to compose our layered multirate control approach and establish our main theoretical guarantee. 

\section{Layered Multirate Control of Constrained Linear Systems}\label{sec:Layered Multirate Control of Constrained Linear Systems}
So far, we have presented the components that constitute our proposed solution to~\Cref{problem}. In particular, we have developed a systematic control design with guaranteed tracking precision and propagated the constraints of the lower-layer system to the higher-layer system.
We are now ready to introduce these components into the layered multirate control architecture depicted in Fig.~\ref{fig:hier_control_architecture}.

Let $\texttt{Planner}:\setR^{\barn}\to\setR^{\barm}$ denote a given planner for $\barSigma$ that satisfies the conditions of~\Cref{prop:Hierarchical Constraint Propagation} and let $u_L(\cdot,\cdot,\cdot)$ be a tracking controller of the form \eqref{eq:tracking_controller}. Recall that the \texttt{Planner} is parametrized by constraint sets $\barX_p$ and $\barU_p$. Moreover, let $\barSigma_H$ be the exact discretization of system $\barSigma$ with zero-order hold inputs \cite{chen1984linear} at the sampling frequency $1/T_H$. Our layered control framework is presented in~\Cref{algorithm}.

\begin{algorithm}[tbh]
\caption{Layered Multirate Control of Constrained Linear Systems}
\label{algorithm}
\textbf{Input:} $x(0)$, $\barx(0)$, $T_L$, $T_H$, $T$, $u_L$, \texttt{Planner}, $\barX_p$, $\barU_p$
\begin{algorithmic}[1]
\State \texttt{$\%$*******Low-frequency planning*******}
\For{$h=0,\ldots,T/T_H-1$}
    \State $\baru_h\gets\texttt{Planner}(\barx(hT_H);\barX_p,\barU_p)$
    \State Apply $\baru_h$ to $\barSigma_H$
    \State \texttt{$\%$******High-frequency tracking*****}
    \For{$\ell=0,\ldots,T_H/T_L-1$}
        \State $u_\ell\gets u_L(\baru_h,\barx(hT_H+\ell T_L),x(hT_H+\ell T_L))$
        \State Apply $u_\ell$ to $\Sigma_L$
    \EndFor
\EndFor
\end{algorithmic}
\end{algorithm}

In the following theorem, we provide our main theoretical result, which follows directly from Lemma~\ref{lem:tracking_precision_guarantee}, Theorem~\ref{theorem:Tracking_using_Discrete-Time_Simulation_Functions}, and Proposition~\ref{prop:Hierarchical Constraint Propagation}.

\begin{theorem}[Layered Multirate Control of Constrained Linear Systems]\label{theorem:main_result}
Consider the layered multirate control architecture depicted in Fig.~\ref{fig:hier_control_architecture}. Let $\texttt{Planner}:\setR^{\barn}\to\setR^{\barm}$ denote a planner parametrized by constraint sets $\barX_p\subseteq\barX$ and $\barU_p\subseteq\barU$ for the state and input of $\barSigma$, respectively. Let the tracking controller $u_L:\setR^{\barm}\times\setR^{\barn}\times\setR^n\to\setR^m$ be as in Theorem~\ref{theorem:Tracking_using_Discrete-Time_Simulation_Functions} and let the tracking precision $\varepsilon$ be as in \eqref{eq:precision}. Assume that $\barX_p$, $\barU_p$, and $Q$ satisfy the conditions of~\Cref{prop:Hierarchical Constraint Propagation}. Fix a pair of initial states $(\barx(0),x(0))\in\calI$ with $\barx(0)\in\barX_p$. Then, by applying the framework described in~\Cref{algorithm}, the closed-loop system $\Sigma$ is ensured to satisfy the tracking guarantee \eqref{eq:precision_guarantee} and the constraints \eqref{eq:safety_guarantee} and \eqref{eq:input_constraint_guarantee2}.
\end{theorem}

We deduce that our framework guarantees the lower-layer system satisfies its output constraints at all high-level time steps, while ensuring its input constraints hold at all low-level time steps. We note that ideas from \cite{barcelliy2010hierarchical} could be employed in combination with our approach to guarantee the satisfaction of $\Sigma$'s output and input constraints for any matrix $Q$ at all low-level time steps (see Remark~\ref{rem:safety_at_low_level_time_steps}).

Beyond providing a systematic framework for layered multirate control of constrained linear systems, our approach guarantees output tracking with a computable precision $\varepsilon$, for all $t=0,T_L,\ldots,T$. Hence, if the closed-loop system $\barSigma_L$ is asymptotically stable, the closed-loop system $\Sigma_L$ is guaranteed to converge to an $\varepsilon$-neighborhood of the origin. Similarly, if $\barSigma_L$ tracks a desired path with precision $\varepsilon'\geq0$, $\Sigma_L$ is ensured to remain within $\varepsilon+\varepsilon'$ of the desired path. An example of motion planning that showcases our results is given in the following section.

\section{Case Study: Navigation of a Mobile Robot}\label{sec:Case Study: Navigation of a Mobile Robot}
\begin{figure}[tbh]
    \centering
    \includegraphics[width=0.65\linewidth]{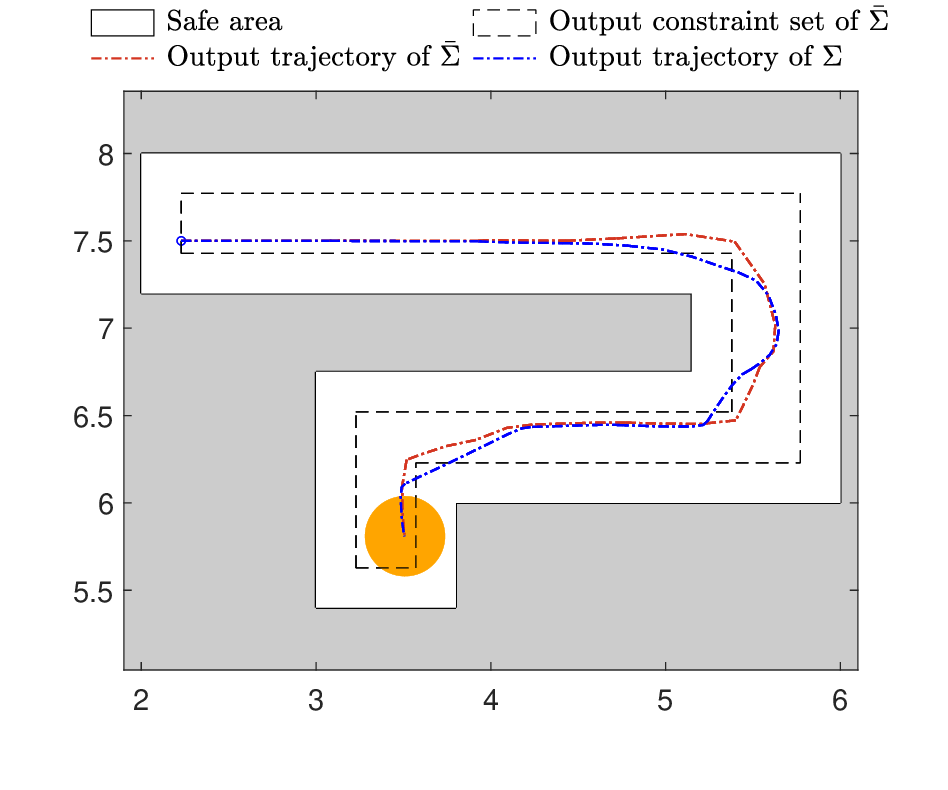}
    \caption{Output trajectories of the lower-layer system $\Sigma$ and the higher-layer system $\barSigma$. The orange disk represents the region that $\Sigma$ aims to reach. The output of $\Sigma$ remains within the safe area at all times, despite temporarily leaving the output constraint set of $\barSigma$ when $\barSigma$ changes direction.}
    \label{fig:maze_plot}
\end{figure}

In this section, we apply our layered multirate control approach to navigate a robot in a maze-like environment. In particular, we consider a planar robot with dynamics model:
\begin{equation*}
    \Sigma:\left\{
            \begin{array}{lcr}
               \dot{p}(t)=v(t) \\
               \dot{v}(t)=a(t) \\
               \dot{a}(t)=u(t) \\
                y(t) = p(t)
            \end{array}
            \right.,
\end{equation*}
where $p(t)\in\setR^2$ is the position, $v(t)\in\setR^2$ is the velocity, and $a(t)\in\setR^2$ is the acceleration. We control the robot via the rate of acceleration,
which is limited by the constraint $u(t)\in[-2,2]^2$. The output constraint set $\calY\subseteq\setR^2$ represents the safe region (white) in Figs.~\ref{fig:failure_plot} and~\ref{fig:maze_plot}.  We perform motion planning with the kinematics model:
\begin{equation*}
    \barSigma:\left\{
            \begin{array}{lcr}
               \dot{\bbar{p}}(t)=\bbar{v}(t) \\
               \dot{\bbar{v}}(t)=\bbar{u}(t) \\
                \bbar{y}(t) = \bbar{p}(t)
            \end{array}
            \right.,
\end{equation*}
where $\bbar{p}(t)\in\setR^2$ is the position, $\bbar{v}(t)\in\setR^2$ is the velocity, and $\bbar{u}(t)\in\setR^2$ is the acceleration, which is employed to control $\barSigma$ and is restricted within $[-1,1]^2$. The state constraint set $\barX$ is given by $\barX=\{(\bbar{p},\bbar{v})\in\setR^4:\bbar{p}\in\calY,\bbar{v}\in\setR^2\}$. The set of initial states for the systems is given by $\calI=\{(p,v,a,\bbar{p},\bbar{v})\in\setR^{10}:p=\bbar{p},p\in\calY,v=0,a=0,\bbar{v}=0\}$. Note that while the outputs of $\Sigma$ and $\barSigma$ share a common physical interpretation (i.e., position), their states and control inputs correspond to different physical quantities. 

Assume that the robot starts at $(2.2,7.5)$ with zero velocity and acceleration. Our goal is to guide it to the disk of radius 0.25, centered at $(3.5,5.7)$, in the environment depicted in Fig.~\ref{fig:maze_plot} (see orange disk). For our mission to be feasible, it is required that the planning constraint sets $\barX_p$ and $\barU_p$ for system $\barSigma$ be nonempty. Since these constraint sets are computed based on the tracking control design, we derive controllers of the form \eqref{eq:tracking_controller} for various sampling frequencies $1/T_L$. For all the sampling frequencies $1/T_L$, we can select a matrix $P$ that leads to $V(\barx(0),x(0))=0$, for all $(\barx(0),x(0))\in\calI$, which implies that $\varepsilon=\gamma\baru_{\max}$ given \eqref{eq:precision}.  

\begin{figure}[tbh]
    \centering
    \includegraphics[width=0.65\linewidth]{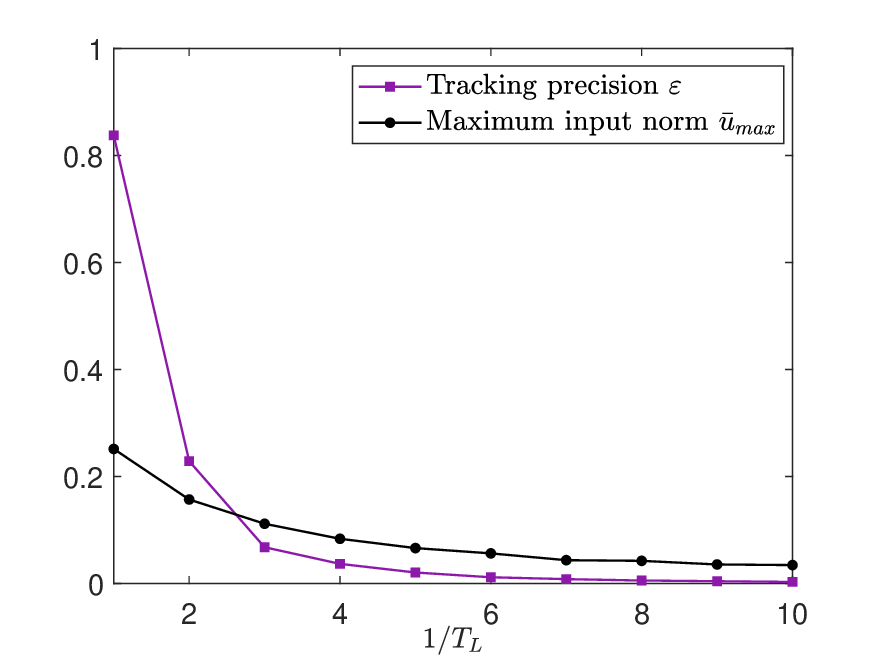}
    \caption{Tracking precision of $\barSigma_L$ by $\Sigma_L$ and maximum input norm for $\barSigma$ across different sampling frequencies $1/T_L$. We observe that higher sampling frequencies improve tracking precision and lead to more restricted inputs for the higher-layer system $\barSigma$.}
    \label{fig:epsilon}
\end{figure}

Fig.~\ref{fig:epsilon} depicts the tracking precision $\varepsilon$ and the maximum allowable input norm $\baru_{\max}$ of $\barSigma$, for $1/T_L=1,2,\ldots,10$ Hz. We observe that higher frequencies $1/T_L$ lead to improved tracking precision. This improvement was expected, as higher values of $1/T_L$ allow $\Sigma$ to update its inputs more frequently, thus enabling it to better track the output of the higher-layer system. Furthermore, increasing $1/T_L$ results in a decrease in the maximum input norm $\baru_{\max}$, which implies reduced control authority for the planner. The decrease in $\baru_{\max}$ can be explained by our computation of the matrix $M$ in \eqref{eq:simulation_function} and the gain $K$ in \eqref{eq:tracking_controller}, which focuses solely on providing a tight precision $\varepsilon$ (see the method in Appendix~\ref{app:Computation of M and K using Semidefinite Programming}). Alternative computational methods, allowing us to attain larger $\baru_{\max}$ while maintaining a tight precision, will be explored in future work.

Observe also in Fig.~\ref{fig:epsilon} that sampling frequencies $1/T_L$ of at least 2 Hz result in a tracking precision $\varepsilon\leq0.23$. Since the minimum width of corridors in our environment is 0.75, for all these values we can compute a nonempty state constraint set $\barX_p$---see the derivation of $\barX_p$ in Appendix~\ref{app:Derivation of the state constraint set barX}.  Furthermore, the initial point as well as a part of the target disk lie within the output constraint set determined by $\barX_p$, which suggests our mission is feasible for all $1/T_L=2,\ldots,10$ Hz. We remark that the output constraint set excludes the initial point that led to safety violations in Fig.~\ref{fig:failure_plot}, rendering it infeasible.\footnote{Fig.~\ref{fig:failure_plot} corresponds to the same pair of systems as Fig.~\ref{fig:maze_plot}, and a different reference path for $\barSigma$, without using though the propagated constraints in the planner.} For the mission of interest, we select $1/T_L=2$ Hz, which ensures feasibility of the mission while allowing for a relatively large $\baru_{\max}$.

We design a reference path from the initial point to the target disk by selecting sequential points within the allowable output set determined by $\barX_p$. For each reference point, our planner employs a model predictive controller \cite{limon2008mpc} for the higher-layer system $\barSigma$, running at a sampling frequency $1/T_H=1\;\text{Hz}$. We note that our ability to choose a higher sampling frequency for the bottom control layer of our architecture is critical to ensuring feasibility of our mission.\footnote{If we had to pick $1/T_L=1/T_H$ (i.e., $1/T_L=1$ Hz), the mission would become infeasible, as the constraint set $\barX_p$ would be empty for the corresponding tracking precision $\varepsilon=0.83$.}

\begin{figure}[tbh]
    \centering
    \includegraphics[width=0.65\linewidth]{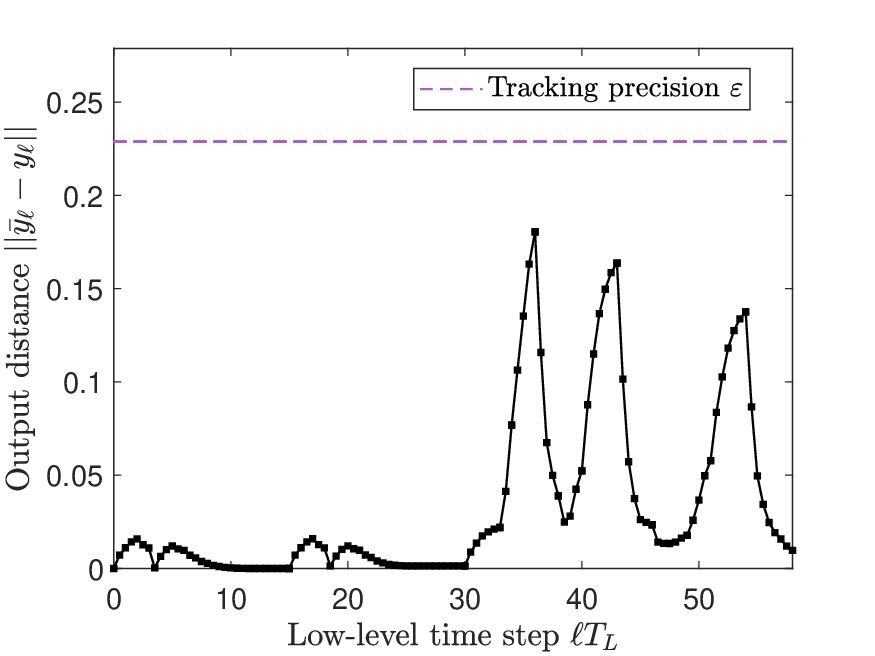}
    \caption{Output distance of $\Sigma$ and $\barSigma$ for the trajectories depicted in Fig.~\ref{fig:maze_plot} at the low-level time steps. Increases of the output distance correspond to time steps where the direction of $\barSigma$ changes. We observe that the tracking precision $\varepsilon$ provides a quite tight higher bound for the output distance.}
    \label{fig:output_distance}
\end{figure}

\begin{figure}[H]
    \centering
    \includegraphics[width=0.7\linewidth]{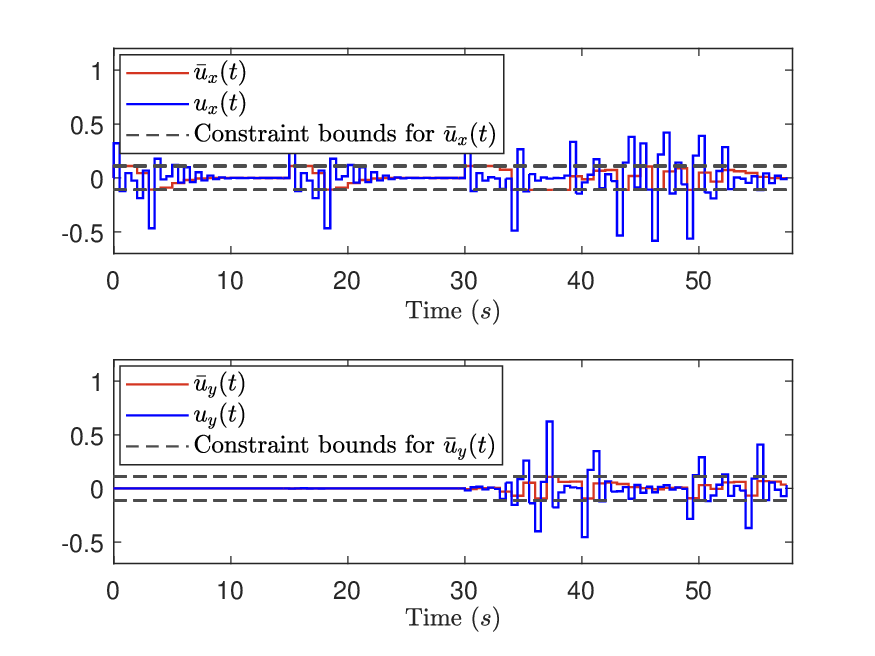}
    \caption{Control input components of $\Sigma$ and $\barSigma$ for the trajectories depicted in Fig.~\ref{fig:maze_plot}. We observe that both input components, $u_x(t)$ and $u_y(t)$, of $\Sigma$ remain within their constraint interval $[-2,2]$.} 
    \label{fig:inputs}
\end{figure}

Fig.~\ref{fig:maze_plot} shows the resulting output trajectories of the lower-layer system $\Sigma$ (in blue) and the higher-layer system $\barSigma$ (in red). We observe that $\Sigma$ remains in the safe area for the entire duration of the mission, verifying our safety guarantee \eqref{eq:safety_guarantee}. Deviation between the two output trajectories is noted at times when the direction of $\barSigma$ changes, owing to the additional dynamics constraint $\dot{a}(t)=u(t)$ of system $\Sigma$. At these times, $\Sigma$ leaves the output constraint set of $\barSigma$ (dashed boundary). However, the output distance of the systems remains smaller than the tracking precision $\varepsilon$, thereby ensuring that $\Sigma$ remains within the safe region (white). Beyond that, in Fig.~\ref{fig:output_distance} we observe that the precision $\varepsilon$ provides a quite tight upper bound for the output distance of the systems during the mission. Finally, Fig.~\ref{fig:inputs} shows the temporal evolution of the input $u(t):=(u_x(t),u_y(t))$ and the input $\baru(t):=(\baru_x(t),\baru_y(t))$ of $\Sigma$ and $\barSigma$, respectively. We can see that the input constraint $u(t)\in[-2,2]^2$ of system $\Sigma$ is satisfied at all times, which verifies the guarantee \eqref{eq:input_constraint_guarantee} of Theorem~\ref{theorem:main_result}.

\section{Discussion and Future Work}
Our paper provides a principled and streamlined approach for layered multirate control of linear systems subject to output and input constraints. In future work, we aim to extend our framework to the more realistic case of nonlinear lower-layer systems, which could have a substantial impact on practical control applications.  

Beyond that, our paper points to several additional research directions. First, in order to allow for more flexibility in planning, we could  derive a method for computing the free parameters of our tracking controller such that the input constraints for the higher-layer system are less conservative. We could also consider designing a robust variant of our approach for systems with disturbances.  Furthermore, additional planning constraints could be enforced to guarantee that the output constraints of the lower-layer system are satisfied at every low-level time step. Finally, analyzing the trade-off between the complexity of the higher-layer system and the computational efficiency of the planner could provide insights into the choice of higher-layer systems. 

\bibliographystyle{IEEEtran} 
\bibliography{arxiv_version}

\appendix
\bigskip
\medskip
\noindent\textbf{\Large Appendix}
\section{Proofs}\label{app:Proofs}
\subsection{Proof of~\Cref{lem:tracking_precision_guarantee}}
We will prove the tracking error bound in \eqref{eq:SF_guarantee} by adapting the proof of \cite[Theorem 1]{Girard2009} to discretized systems. Fix any $(\barx_0,x_0)\in\calI$ and let us define:
\begin{equation*}
    \alpha=\max\left\{\sqrt{V(\barx_0,x_0)},\gamma\baru_{\max}\right\}.    
\end{equation*}
By definition of $\alpha$, we have $\sqrt{V(\barx_0,x_0)}\leq\alpha$. Assume there exists $\ell_1\in\setN_+$ such that $\sqrt{V(\barx_{\ell_1},x_{\ell_1})}>\alpha$.
Then, there exists $\ell_0\in\{0,\ldots,\ell_1-1\}$ such that $\sqrt{V(\barx_{\ell_0},x_{\ell_0})}\leq\alpha$ and $\sqrt{V(\barx_{\ell},x_\ell)}>\alpha$, for all $\ell\in\{\ell_0+1,\ldots,\ell_1\}$. Then, by the fact that $\baru_\ell\in\barU_p$, with $\max_{\baru\in\calU_p}\norm{\baru}\leq\baru_{\max}$, and the definition of $\alpha$, we have: 
\[
\gamma\norm{\baru_\ell}\leq\gamma\baru_{\max}\leq\alpha<\sqrt{V(\barx_{\ell},x_\ell)},
\]
for all $\ell\in\{\ell_0+1,\ldots,\ell_1\}$. Hence, from inequality \eqref{eq:SF_condition_2} we can conclude that:
\begin{equation*}
    V(\barx_{\ell+1},x_{\ell+1})-V(\barx_\ell,x_\ell)<0,
\end{equation*}
for all $\ell\in\{\ell_0+1,\ldots,\ell_1\}$, which implies that:
\begin{equation}\label{lem:tracking_precision_guarantee_proof_1}
    \sqrt{V(\barx_{\ell_1},x_{\ell_1})}<\sqrt{V(\barx_{\ell_0},x_{\ell_0})}\leq\alpha.
\end{equation}
Inequality \eqref{lem:tracking_precision_guarantee_proof_1} contradicts the assumption that  $\sqrt{V(\barx_{\ell_1},x_{\ell_1})}>\alpha$, and hence we deduce that $\sqrt{V(\barx_{\ell},x_{\ell})}\leq\alpha$, for all $\ell\in\setN$. Employing inequality \eqref{eq:SF_condition_1}, we obtain:
\begin{equation*}
    \norm{\bary_\ell-y_\ell}=\norm{\barC\barx_\ell-Cx_\ell}\leq \sqrt{V(\barx_\ell,x_\ell)}\leq\alpha,
\end{equation*}
for all $\ell\in\setN$, which completes the proof. 

$\qed$

\subsection{Proof of~\Cref{lem:M_K_conditions}}
We will prove the lemma by adapting the proof of \cite[Proposition 3.3]{girard2005approximate} to discrete-time systems. By Assumption~\ref{ass:P_conditions}, we know that system $\Sigma_L$ is stabilizable and thus there exists a gain matrix $K\in\setR^{m\times n}$ such that all the eigenvalues of $A_{\textup{cl}}:=A_L+B_LK$ lie strictly within the unit disk. Let $M=N+C^{\intercal}C$, where $N\in\setR^{n\times n}$ is a positive definite matrix to be determined later on. Given that $N$ is positive definite, we can directly deduce that $M$ is positive definite and inequality \eqref{eq:M_condition} is satisfied. We set:
\[
    A_{\textup{cl},\lambda} = \frac{1}{\sqrt{1-2\lambda}} A_{\textup{cl}},
\]
where $\lambda\in(0,1/2)$ is such that the eigenvalues of $A_{\textup{cl},\lambda}$ lie strictly within the unit disk. We note that we can always select a scalar $\lambda$ small enough such that this condition holds. By definition of $M$ and $A_{\textup{cl},\lambda}$, inequality \eqref{eq:M_K_condition} can be equivalently written as:
\begin{align}
    &A_{\textup{cl}}^{\intercal}MA_{\textup{cl}}-(1-2\lambda)M\preceq0 \nonumber\\
    \iff &A_{\textup{cl},\lambda}^{\intercal}MA_{\textup{cl},\lambda}-M\preceq0\nonumber\\
    \label{lem:tracking_precision_guarantee_proof_2}
    \iff &A_{\textup{cl},\lambda}^{\intercal}NA_{\textup{cl},\lambda}-N\preceq-A_{\textup{cl},\lambda}^{\intercal}C^{\intercal}CA_{\textup{cl},\lambda}+C^{\intercal}C.
\end{align}
Let $Q\succeq0$ be such that:
\begin{equation}\label{lem:tracking_precision_guarantee_proof_3}
    A_{\textup{cl},\lambda}^{\intercal}C^{\intercal}CA_{\textup{cl},\lambda}-C^{\intercal}C\preceq Q.
\end{equation}
Then, we can define $N$ as the unique positive definite solution of the Lyapunov equation:
\begin{equation}\label{lem:tracking_precision_guarantee_proof_4}
    A_{\textup{cl},\lambda}^{\intercal}NA_{\textup{cl},\lambda}-N=-Q.
\end{equation}
Substituting \eqref{lem:tracking_precision_guarantee_proof_4} into \eqref{lem:tracking_precision_guarantee_proof_3}, we obtain inequality \eqref{lem:tracking_precision_guarantee_proof_2}. Since \eqref{lem:tracking_precision_guarantee_proof_2} is equivalent to \eqref{eq:M_K_condition}, this completes the proof.

\subsection{Proof of~\Cref{theorem:Tracking_using_Discrete-Time_Simulation_Functions}}
Fix any $\barx_\ell\in\setR^{\barn}$ and $x_\ell\in\setR^n$. Employing \eqref{eq:simulation_function}, \eqref{eq:M_condition}, and \eqref{eq:P_condition}, we obtain:
\begin{align*}
    V(\barx_\ell,x_\ell)=(x_\ell-P\barx_\ell)^{\intercal}M(x_\ell-P\barx_\ell)\geq(x_\ell-P\barx_\ell)^{\intercal}C^{\intercal}C(x_\ell-P\barx_\ell)=\norm{\barC\barx_\ell-Cx_\ell}^2,
\end{align*}
which implies that condition \eqref{eq:SF_condition_1} is satisfied. Let $\barU_p$ be a compact set and let $\baru_{\max}>0$ be such that $\max_{\baru\in\barU_p}\norm{\baru}\leq\baru_{\max}$. Moreover, set $\Acl=A_L+B_LK$ and let $\baru_\ell\in\barU_p$ be such that $\gamma^2\norm{\baru_\ell}^2<\sqrt{V(\barx_\ell,x_\ell)}$. Given the expression of $\barx_{\ell+1}$ and $x_{\ell+1}$ in~\Cref{def:Discrete-Time Simulation Functions and Associated Tracking Controllers}, we can write:
\begin{align*}
 &x_{\ell+1}-P\barx_{\ell+1} \\
 =\; &A_Lx_\ell+B_Lu_L(\baru_\ell,\barx_\ell,x_\ell)-P(\barA_L\barx_\ell+\barB_L\baru_\ell)\\
 =\; &A_Lx_\ell+B_L(R\baru_\ell+Q\barx_\ell+K(x_\ell-P\barx_\ell))-P(\barA_L\barx_\ell+\barB_L\baru_\ell)\hspace*{1cm}\text{(from \eqref{eq:tracking_controller})}\\
 =\; &\Acl(x_\ell-P\barx_\ell)+(A_LP+B_LQ-P\barA_L)\barx_\ell+(B_LR-P\barB_L)\baru_\ell \\
    =\; &\Acl(x_\ell-P\barx_\ell)+(B_LR-P\barB_L)\baru_\ell.\hspace*{5.15cm}\text{(from \eqref{eq:P_Q_condition})}
\end{align*}
Then, by performing straightforward algebraic manipulations, we get:
\begin{align}
    &V(\barx_{\ell+1},x_{\ell+1})-V(\barx_\ell,x_\ell) \nonumber\\
    =\; &(x_{\ell+1}-P\barx_{\ell+1})^{\intercal}M(x_{\ell+1}-P\barx_{\ell+1})-(x_\ell-P\barx_\ell)^{\intercal}M(x_\ell-P\barx_\ell) \nonumber\\
    =\; &\left(\Acl(x_\ell-P\barx_\ell)+(B_LR-P\barB_L)\baru_\ell\right)^{\intercal}M\left(\Acl(x_\ell-P\barx_\ell)+(B_LR-P\barB_L)\baru_\ell\right)\nonumber\\
    &-(x_\ell-P\barx_\ell)^{\intercal}M(x_\ell-P\barx_\ell) \nonumber\\
    =\; &(x_\ell-P\barx_\ell)^{\intercal}(\Acl^{\intercal}M\Acl-M)(x_\ell-P\barx_\ell)+\norm{M^{1/2}(B_LR-P\barB_L)\baru_\ell}^2\nonumber\\
    &+2(x_\ell-P\barx_\ell)^{\intercal}\Acl^{\intercal}M(B_LR-P\barB_L)\baru_\ell\nonumber\\
    \leq\; &(x_\ell-P\barx_\ell)^{\intercal}(\Acl^{\intercal}M\Acl-M)(x_\ell-P\barx_\ell)+\norm{M^{1/2}(B_LR-P\barB_L)\baru_\ell}^2\nonumber\\
    \label{thm1:proof_1}
    &+2\norm{M^{1/2}\Acl(x_\ell-P\barx_\ell)}\norm{M^{1/2}(B_LR-P\barB_L)\baru_\ell},
\end{align}
where the last step follows from a direct application of Cauchy-Schwarz inequality. By setting:
\[
\alpha=\frac{1}{\sqrt{\mu}}\norm{M^{1/2}\Acl(x_\ell-P\barx_\ell)},\; \beta=\sqrt{\mu}\norm{M^{1/2}(B_LR-P\barB_L)\baru_\ell}
\]
where $\mu=(1-\lambda)/\lambda$, and using the basic inequality:
\[
(\alpha-\beta)^2\geq0 \iff 2\alpha\beta\leq\alpha^2+\beta^2,
\]
we can deduce that:
\begin{align}
    &2\norm{M^{1/2}\Acl(x_\ell-P\barx_\ell)}\norm{M^{1/2}(B_LR-P\barB_L)\baru_\ell}\nonumber\\
    =\; &2\alpha\beta\nonumber\\
    \leq\; &\alpha^2+\beta^2\nonumber\\
    =\; &\frac{1}{\mu}\norm{M^{1/2}\Acl(x_\ell-P\barx_\ell)}^2+\mu\norm{M^{1/2}(B_LR-P\barB_L)\baru_\ell}^2\nonumber\\
    \label{thm1:proof_2}
    =\; &\frac{1}{\mu}(x_\ell-P\barx_\ell)^{\intercal}\Acl^{\intercal} M\Acl(x_\ell-P\barx_\ell)+\mu\norm{M^{1/2}(B_LR-P\barB_L)\baru_\ell}^2.
\end{align}
Employing \eqref{thm1:proof_2}, inequality \eqref{thm1:proof_1} yields:
\begin{align}
    &V(\barx_{\ell+1},x_{\ell+1})-V(\barx_\ell,x_\ell)\nonumber\\
    \leq\; &(x_\ell-P\barx_\ell)^{\intercal}\left(\left(1+\frac{1}{\mu}\right)\Acl^{\intercal}M\Acl-M\right)(x_\ell-P\barx_\ell)+(1+\mu)\norm{M^{1/2}(B_LR-P\barB_L)\baru_\ell}^2\nonumber\\
    \label{thm1:proof_3}
    =\; &(x_\ell-P\barx_\ell)^{\intercal}\left(\left(\frac{1}{1-\lambda}\right)\Acl^{\intercal}M\Acl-M\right)(x_\ell-P\barx_\ell)+\frac{1}{\lambda}\norm{M^{1/2}(B_LR-P\barB_L)\baru_\ell}^2.
\end{align}
Given that $\lambda\in(0,1/2)$, notice that \eqref{eq:M_K_condition} implies that:
\begin{align}
    &\Acl^{\intercal}M\Acl-M\preceq-2\lambda M \nonumber\\
    &\Acl^{\intercal}M\Acl-(1-\lambda)M\preceq-\lambda M \nonumber\\
    \label{thm1:proof_4}
    &\left(\frac{1}{1-\lambda}\right)\Acl^{\intercal}M\Acl-M\preceq-\frac{\lambda}{1-\lambda}M.
\end{align}
Let $\gamma$ be defined as in the theorem statement. Using \eqref{thm1:proof_4}, Cauchy-Schwarz inequality, and \eqref{eq:simulation_function}, relationship \eqref{thm1:proof_3} yields:
\begin{align*}
    &V(\barx_{\ell+1},x_{\ell+1})-V(\barx_\ell,x_\ell)\nonumber\\
    \leq\; &-\left(\frac{\lambda}{1-\lambda}\right)(x_\ell-P\barx_\ell)^{\intercal}M(x_\ell-P\barx_\ell)+\frac{1}{\lambda}\norm{M^{1/2}(B_LR-P\barB_L)}^2\norm{\baru_\ell}^2\nonumber\\
    =\; &-\left(\frac{\lambda}{1-\lambda}\right)(V(\barx_\ell,x_\ell)-\gamma^2\norm{\baru_\ell}^2)\\
    <\; &0,
\end{align*}
where the last inequality follows from the fact that $\lambda\in(0,1/2)$ and the assumption that $V(\barx_\ell,x_\ell)<\gamma^2\norm{\baru_\ell}^2$. Hence, we have shown condition \eqref{eq:SF_condition_2} of~\Cref{def:Discrete-Time Simulation Functions and Associated Tracking Controllers} is also satisfied, which completes the proof.

$\qed$

\subsection{Proof of~\Cref{prop:Hierarchical Constraint Propagation}}
Consider the planning constraints:
\begin{align}
    \label{prop1:proof_1}
    &\barx(t)\in\barX_p, \\
    \label{prop1:proof_2}
    &\baru(t)\in\barU_p, 
\end{align}
for all $t=0,T_H,\ldots,T$, where the constraint sets $\barX_p$ and $\barU_p$ satisfy the assumptions stated in the proposition. From \eqref{eq:constraints_condition_1} and \eqref{prop1:proof_1} we obtain:
\begin{align}
    \label{prop1:proof_3}
    \bary(t) = \barC\barx(t)\in\barC\barX_p\subseteq\calY-\varepsilon\calS^p,
\end{align}
for all $t=0,T_H,\ldots,T$. Moreover, from~\Cref{lem:tracking_precision_guarantee} and the definition of $\varepsilon$ in \eqref{eq:precision}, we can conclude that the tracking controller $u_L(\cdot,\cdot,\cdot)$ guarantees bounded tracking error as follows:
\begin{align}
    \label{prop1:proof_4}
    \norm{\bary(t)-y(t)}\leq\varepsilon,\; t=0,T_L,\ldots,T,
\end{align}
for any planner and any pair of initial states $(\barx(0),x(0))\in\calI$ with $\barx(0)\in\barX_p$. From \eqref{prop1:proof_3} and \eqref{prop1:proof_4} we deduce that the closed-loop system $\Sigma$ satisfies the safety guarantee \eqref{eq:safety_guarantee}. 

Equation \eqref{eq:tracking_controller} implies that:
\begin{align}
    \label{prop1:proof_5}
    u(t) &= R\baru(t)+Q\barx(t)+K(x(t)-P\barx(t)) = R\baru(t)+Q\barx(t)+KM^{-1/2}M^{1/2}(x(t)-P\barx(t)),
\end{align}
for all $t=0,T_L,\ldots,T$. Note that the exact discretization $\barSigma_H$ of system $\barSigma$ with zero-order hold inputs implies that $\baru(t)=\baru(hT_H)$, for all $t\in[hT_H,(h+1)T_H)$, $h=0,\ldots,T/T_H-1$. Hence, from \eqref{prop1:proof_2} we can conclude that:
\begin{align}
    \label{prop1:proof_6}
    &\baru(t)\in\barU_p,\; t=0,T_L,\ldots,T.
\end{align}
Furthermore, from \eqref{eq:simulation_function} and the proof of~\Cref{lem:tracking_precision_guarantee} we can directly conclude that:
\begin{align}
    \label{prop1:proof_7}
    \norm{M^{1/2}(x(t)-P\barx(t))}=V(\barx(t),x(t))\leq\max\left\{\sqrt{V(\barx(0),x(0))},\gamma\baru_{\max}\right\}\leq\varepsilon,
\end{align}
for all $t=0,T_L,\ldots,T$. Employing \eqref{eq:constraints_condition_2}, \eqref{prop1:proof_1}, \eqref{prop1:proof_2}, and \eqref{prop1:proof_7}, equation \eqref{prop1:proof_5} directly leads to the guarantee \eqref{eq:input_constraint_guarantee}. If $Q=0$, from \eqref{eq:constraints_condition_2}, \eqref{prop1:proof_5}, \eqref{prop1:proof_6}, and \eqref{prop1:proof_7} we obtain the guarantee \eqref{eq:input_constraint_guarantee2}, thus completing the proof.

$\qed$

\subsection{Propagation of Polytopic Constraints}\label{app:Layered Propagation of Polytopic Constraints}
Consider polyhedral constraint sets for the lower-layer system, given by: 
\begin{align*}
    &\calY=\left\{y\in\setR^p: F_yy\leq f_y\right\},\\
    &\calU=\left\{u\in\setR^m: F_uu\leq f_u\right\},
\end{align*}
as described in Section~\ref{sec:Hierarchical Constraint Propagation}. Let $\barX_p$ and $\barU_p$ be defined as in \eqref{eq:barX_polyhedral} and \eqref{eq:barU_polyhedral}, respectively. Fix any $\barx\in\barX_p$ and $\baru\in\barU_p$. Moreover, let $f_{y,j}$ and $f_{u,j}$ denote the $j$-th element of $f_y$ and $f_u$, respectively. Then, by definition of $\barX_p$ and $\barU_p$, we have:
\begin{align}
    \label{propagation:proof_1}
    F_{y,j}\barC\barx&\leq f_{y,j}-\varepsilon\norm{F_{y,j}},\;\forall j=1,\ldots,d_y,\\
    \label{propagation:proof_2}
    \norm{Q\barx}_{\infty}&\leq\delta,\\
    \label{propagation:proof_3}
    F_{u,j}R\baru&\leq f_{u,j}-\delta\norm{F_{u,j}}_1-\varepsilon\norm{G_{u,j}},\;\forall j=1,\ldots,d_u.
\end{align}
Using \eqref{propagation:proof_1} and Cauchy-Schwarz inequality, we can write:
\begin{align}
    &F_{y,j}\barC\barx\leq f_{y,j}-\norm{e}\norm{F_{y,j}},\;\forall j=1,\ldots,d_y, e\in\varepsilon\calS^p\nonumber\\
    \implies\;  &F_{y,j}\barC\barx\leq f_{y,j}-F_{y,j}e,\;\forall j=1,\ldots,d_y, e\in\varepsilon\calS^p\nonumber\\
    \iff\; &F_{y,j}(\barC\barx+e)\leq f_{y,j},\;\forall j=1,\ldots,d_y, e\in\varepsilon\calS^p\nonumber\\
    \iff\; &\barC\barx+\varepsilon\calS^p\subseteq\calY \nonumber\\
    \iff\; &\barC\barx\in\calY-\varepsilon\calS^p,\nonumber
\end{align}
which implies that condition \eqref{eq:constraints_condition_1} holds. From \eqref{propagation:proof_2}, \eqref{propagation:proof_3}, and Cauchy-Schwarz inequality, we obtain:
\begin{align}
    &F_{u,j}R\baru\leq f_{u,j}-\norm{Q\barx}_{\infty}\norm{F_{u,j}}_1-\norm{e}\norm{G_{u,j}},\;\forall j=1,\ldots,d_y, \barx\in\barX_p, e\in\varepsilon\calS^n\nonumber\\
    \implies\; &F_{u,j}R\baru\leq f_{u,j}-F_{u,j}Q\barx-G_{u,j}e,\;\forall j=1,\ldots,d_y, \barx\in\barX_p, e\in\varepsilon\calS^n\nonumber\\
    \iff\; &F_{u,j}(R\baru+Q\barx+KM^{-1/2}e)\leq f_{u,j},\;\forall j=1,\ldots,d_y, \barx\in\barX_p, e\in\varepsilon\calS^n\hspace{.7cm}(G_{u,j}=F_{u,j}KM^{-1/2})\nonumber\\
    \iff\; &R\baru+Q\barX_p+\varepsilon KM^{-1/2}\calS^n\subseteq\calU \nonumber\\
    \iff\; &R\baru+Q\barX_p\subseteq\calU-\varepsilon KM^{-1/2}\calS^n,\nonumber
\end{align}
which implies that condition \eqref{eq:constraints_condition_2} holds.

$\qed$

\section{Computation of the Matrices $M$ and $K$ using Semidefinite Programming}\label{app:Computation of M and K using Semidefinite Programming}
For a given $\lambda\in(0,1/2)$, we can compute the matrices $M$ and $K$ by setting $\tilde{M}=M^{-1}$ and $\tilde{K}=KM^{-1}$, and solving the following optimization problem:
\begin{subequations}\label{optim_problem}
\begin{align}
    \max_{\tilde{M}\succ0,\, \tilde{K}}\;&\lambda_{\min}(\tilde{M}) \\
    \subto  \label{eq:LMI_constraint_2}
            &\begin{bmatrix}
                I & C\tilde{M} \\
                \tilde{M}C^{\intercal} & \tilde{M}             
            \end{bmatrix}\succeq0\\
            \label{eq:LMI_constraint_3}
            &\begin{bmatrix}
                \tilde{M} & A_L\tilde{M}+B_L\tilde{K}\\
                (A_L\tilde{M}+B_L\tilde{K})^{\intercal} & (1-2\lambda)\tilde{M}
            \end{bmatrix}\succeq0,
\end{align}
\end{subequations}
where $\lambda_{\min}(\tilde{M})$ denotes the minimum eigenvalue of $\tilde{M}$. Using Schur complements, it is straightforward to show that the linear matrix inequality (LMI) constraints \eqref{eq:LMI_constraint_2} and \eqref{eq:LMI_constraint_3} are equivalent to \eqref{eq:M_condition} and \eqref{eq:M_K_condition}, respectively. \Cref{lem:tracking_precision_guarantee} guarantees that the above problem is feasible for some $\lambda\in(0,1/2)$. Maximizing $\lambda_{\min}(\tilde{M})$ is equivalent to minimizing the maximum eigenvalue of $M$, which results in a small value for the first component of the maximum in the tracking precision \eqref{eq:precision} (see \eqref{eq:simulation_function}). In the case study of Section~\ref{sec:Case Study: Navigation of a Mobile Robot}, we observe that the optimization cost of \eqref{optim_problem} is critical to obtaining a tight tracking precision (see Fig.~\ref{fig:output_distance}). Problem \eqref{optim_problem} can be easily solved by considering its epigraph form \cite{boyd2004convex}:
\begin{subequations}
\label{optim_problem_2}
\begin{align}
    \max_{\tilde{M},\, \tilde{K},\, s}\;&s \\
    \subto  \label{eq:LMI_constraint2_0}
            &s\geq \epsilon \\
            \label{eq:LMI_constraint2_1}
            &\tilde{M}\succeq sI \\
            \label{eq:LMI_constraint2_2}                        &\begin{bmatrix}
                I & C\tilde{M} \\
                \tilde{M}C^{\intercal} & \tilde{M}             
            \end{bmatrix}\succeq0\\
            \label{eq:LMI_constraint2_3}            &\begin{bmatrix}
                \tilde{M} & A_L\tilde{M}+B_L\tilde{K}\\
                (A_L\tilde{M}+B_L\tilde{K})^{\intercal} & (1-2\lambda)\tilde{M}
            \end{bmatrix}\succeq0,
\end{align}
\end{subequations}
where $\epsilon$ is a small positive constant. The inequality constraint \eqref{eq:LMI_constraint2_0} along with the LMI constraint \eqref{eq:LMI_constraint2_1} guarantees that $M$ is positive definite (due to the use of $\epsilon>0$). The LMI constraint \eqref{eq:LMI_constraint2_1} along with the cost function $s$ ensure the maximization of $\lambda_{\min}(\tilde{M})$. The LMI constraints \eqref{eq:LMI_constraint2_2} and \eqref{eq:LMI_constraint2_3} are the same as the LMI constraints \eqref{eq:LMI_constraint_2} and \eqref{eq:LMI_constraint_3} in problem \eqref{optim_problem}. We note that problem \eqref{optim_problem_2} is a semidefinite program and thus can be solved efficiently using standard solvers, such as those in CVX \cite{grant2014cvx}.

\section{Derivation of the state constraint set $\bar{\mathcal{X}}_p$ in Section~\ref{sec:Case Study: Navigation of a Mobile Robot}}\label{app:Derivation of the state constraint set barX}

\begin{figure}[tbh]
    \centering
    \includegraphics[width=0.7\linewidth]{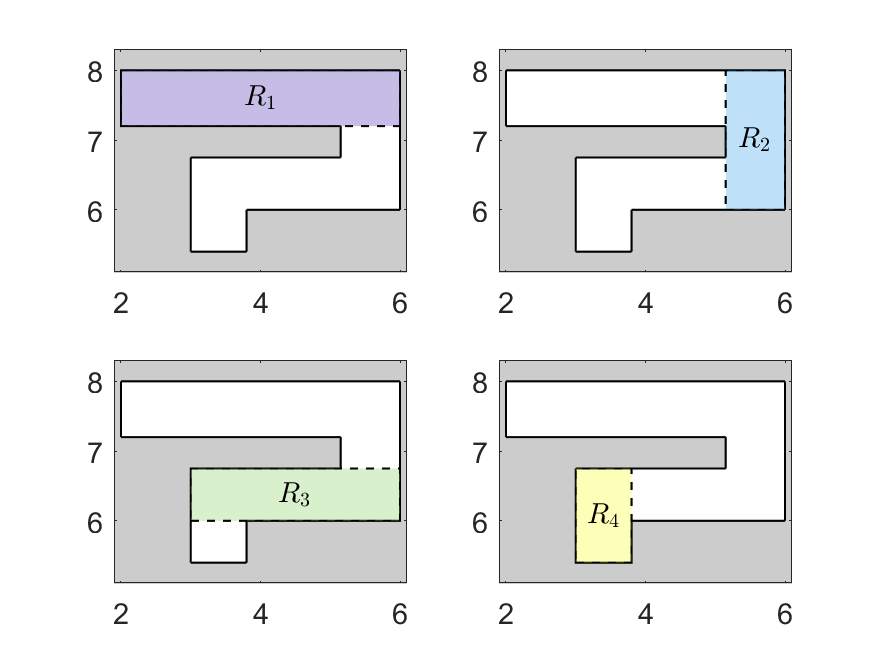}
    \caption{The union of the rectangles $R_1$, $R_2$, $R_3$, and $R_4$ constitutes the nonconvex safe set $\calY$ of Fig.~\ref{fig:maze_plot}. Constraint propagation from the lower- to the higher-layer system relies sequentially on the rectangles $R_1$, $R_2$, $R_3$, and $R_4$, as the robot moves from one rectangle to the next.}
    \label{fig:rectangles}
\end{figure}

In this section, we briefly explain the derivation of the planning constraint set $\barX_p$ for the nonconvex safe region considered in Section~\ref{sec:Case Study: Navigation of a Mobile Robot}. Following the procedure outlined in Remark~\ref{rem:derivation_of_planning_constraint_sets}, we can consider the safe set in Fig.~\ref{fig:maze_plot} as the union of the rectangles $R_i$ depicted in Fig.~\ref{fig:rectangles}. Constraint propagation for each of the rectangles can be performed as described in Section~\ref{sec:Hierarchical Constraint Propagation}. A standard linear model predictive controller \cite{limon2008mpc} can handle the resulting polyhedral constraint sets \eqref{eq:barX_polyhedral} and \eqref{eq:barU_polyhedral}.

\end{document}